\documentclass[onecolumn]{IEEEtran}
\usepackage{cite}
\usepackage{amsmath,amssymb,amsfonts}
\usepackage{algorithmic}
\usepackage{graphicx}
\usepackage{algorithm,algorithmic}
\usepackage[hidelinks]{hyperref}
\usepackage{textcomp}
\usepackage{booktabs}
\newcommand{\codeURL}{\url{https://github.com/qxyzs/SDIP}}

\begin{document}
\title{A Zero-Shot Deep Image Prior Framework for Denoising and Deconvolution in Fluorescence Microscopy}

\author{
Xiangyu Qian$^{1,2}$,
Jing Liu$^{1}$,
Yunqing Tang$^{1,\ast}$,
Luru Dai$^{1,2,\ast}$,
Qiushi Li$^{1,\ast}$

$^{1}$ Wenzhou Key Laboratory of Biomedical Imaging, Zhejiang Key Laboratory of Soft Matter Biomedical Materials, Wenzhou Institute, University of Chinese Academy of Sciences, Wenzhou, Zhejiang 325000, China

$^{2}$ Postgraduate Training Base Alliance of Wenzhou Medical University, Wenzhou, Zhejiang 325035, China

\emph{E-mails:tangyunqing@wiucas.ac.cn; dai@wiucas.ac.cn; liqs@wiucas.ac.cn}

}

\maketitle
\begin{abstract}
Fluorescence microscopy images are degraded by noise and diffraction-induced blur, which compromise structural fidelity and limit quantitative analysis. Supervised deep learning methods achieve impressive restoration performance but require large-scale paired datasets that are difficult to obtain in practice. To address this issue, we propose SDIP, a zero-shot deep image prior (DIP) framework that sequentially performs denoising and deconvolution without external training data. An aSeqDIP-based module first suppresses noise while preserving fine structures through sequential autoencoding regularization. In the deconvolution stage, a wavelet-based background correction step is incorporated before the proposed RLG-DIP module performs artifact-reduced deconvolution. RLG-DIP uses the Richardson–Lucy deconvolution result as a physically consistent guidance prior, integrating the imaging model with the implicit prior of DIP to stabilize the ill-posed deconvolution process. Experiments on the BioSR dataset across multiple cellular structures demonstrate that SDIP improves both signal-to-noise ratio and resolution, achieving superior visual quality and improved quantitative performance on most evaluated structures. The proposed framework may also provide useful insights for designing physically guided DIP methods for other inverse problems.
\end{abstract}


\section{Introduction}
	
	Fluorescence microscopy has become an essential tool in modern life sciences for visualizing cellular morphology, subcellular organization, and dynamic biological processes\cite{schermelleh2019super}. Despite its widespread use, image quality is fundamentally constrained by both optical and experimental factors. In practice, photobleaching and phototoxicity restrict imaging conditions to low excitation intensity and short exposure times, resulting in a limited photon budget, increased noise, and a reduced signal-to-noise ratio (SNR). Meanwhile, the optical imaging process, governed by the point spread function (PSF), inherently introduces blur\cite{zhang2019poisson}. Together, these factors lead to degraded observations that impair structural fidelity and limit the reliability of downstream quantitative analysis\cite{bertero_introduction_2021}. These challenges have driven the development of computational image restoration methods, particularly denoising and deconvolution approaches\cite{tian2025computational}.
	
	Existing computational restoration methods can be broadly categorized into analytical and learning-based approaches\cite{tian2025computational}. Early analytical methods explicitly model the imaging process using the PSF and noise statistics, offering strong interpretability and physical consistency\cite{richardson1972bayesian},\cite{lucy1974iterative}. However, their performance often depends on simplified assumptions and careful parameter tuning, which limits robustness under low SNR conditions\cite{bertero_introduction_2021}. More recently, deep learning-based methods have achieved remarkable performance by learning data-driven mappings from degraded observations to high-quality images. In particular, supervised approaches have demonstrated state-of-the-art results in denoising and deconvolution-related restoration tasks\cite{qiao2021evaluation},\cite{li2022incorporating},\cite{qiao2023rationalized}. However, their success relies heavily on large-scale paired training datasets with high-quality ground truth, which are often difficult to obtain in fluorescence microscopy.
	
	To mitigate the reliance on clean ground-truth data, increasing attention has been devoted to label-free image restoration methods in fluorescence microscopy. A major line of work follows the Noise2Noise paradigm \cite{lehtinen2018noise2noise}  and its self-supervised variant\cite{krull2019noise2void},\cite{batson2019noise2self},\cite{quan2020self2self},\cite{lequyer2022fast},\cite{chen2024self},\cite{qiao2024zero},\cite{qu2024self}, which exploit temporal, spatial, or statistical redundancy in the observed data to construct surrogate supervision. Although effective in denoising, reconstruction, and resolution enhancement, these methods usually rely on specific assumptions about data redundancy and noise statistics. In contrast, Deep Image Prior (DIP) \cite{ulyanov2018deep}  offers a fundamentally different zero-shot strategy by restoring images through single-image optimization with the implicit bias of convolutional networks, without requiring surrogate image pairs. This makes DIP particularly appealing for fluorescence microscopy, where clean references are often unavailable. However, directly applying DIP remains challenging because the optimization can gradually fit noise, and its generic implicit prior lacks explicit guidance from the microscopy imaging model.
	
	Motivated by the above challenges, we propose a DIP-based zero-shot restoration method for fluorescence microscopy image enhancement. The proposed approach extends DIP to fluorescence microscopy through a stage-wise restoration strategy, which decomposes the restoration process into denoising and deconvolution stages to better address coupled noise and optical blur degradations. Such a staged design is adopted because directly performing deconvolution on noisy fluorescence images tends to amplify noise and destabilize the restoration process. For the deconvolution stage, an RL-guided DIP (RLG-DIP) method is further introduced. This deconvolution strategy is motivated by the fact that RL-based deconvolution is effective for blur removal but highly sensitive to noise, whereas DIP tends to favor cleaner image structures during optimization. By introducing the RL deconvolution result as a guidance constraint in DIP optimization, the method combines RL-based structural guidance with the implicit prior of DIP, suppressing artifacts and enhancing structural fidelity. Extensive experiments on real fluorescence microscopy data demonstrate that the proposed method achieves improved image quality and more reliable structural recovery.
	
	The main contributions of this work are summarized as follows:
	
	(1) A stage-wise DIP-based restoration framework (SDIP) for fluorescence microscopy, which decomposes the restoration process into denoising and deconvolution stages to address coupled noise and optical blur degradations.
	
	(2) An RL-guided DIP (RLG-DIP) method for deconvolution, which introduces RL-based structural guidance into DIP optimization, enabling effective artifact suppression and improved structural fidelity. It also suggests that the proposed adaptation of DIP for fluorescence microscopy may inform DIP design for other restoration tasks.
	
	(3) Extensive experiments on real fluorescence microscopy data demonstrate improved image quality and more reliable structural recovery.

	\section{RELATED WORK}
	\subsection{Denoising in fluorescence microscopy}
	
	Early fluorescence microscopy denoising methods mainly rely on analytical models and hand-crafted priors, such as BM3D\cite{dabov2007image} and statistical noise modeling approaches\cite{mandracchia_fast_2020}. These methods are interpretable and training-free, but their performance is often limited under low SNR conditions and depends on accurate noise assumptions.
	
	Deep learning-based denoising methods,especially supervised approaches, have achieved strong restoration performance\cite{weigert2018content}. However, they usually require large-scale paired noisy-clean datasets, which are difficult to acquire in fluorescence microscopy. Self-supervised methods alleviate this requirement by exploiting spatial, temporal, or statistical redundancy\cite{lequyer2022fast}. Nevertheless, these methods often rely on assumptions such as noise independence or masking strategies, and are mainly designed for denoising rather than joint restoration involving both noise and blur.
	
	\subsection{Deconvolution in fluorescence microscopy}
	Deconvolution aims to recover high-resolution structures from PSF-induced blur. Classical methods such as RL deconvolution\cite{richardson1972bayesian,lucy1974iterative} explicitly model the imaging process and are widely used in practice. However, deconvolution is an ill-posed inverse problem and is highly sensitive to noise, often leading to artifact amplification in low-SNR conditions\cite{bertero_introduction_2021}.
	
	Regularization-based methods introduce priors such as sparsity\cite{zhao2022sparse}, smoothness\cite{huang_fast_2018}, or total variation\cite{rudin1992nonlinear} to stabilize reconstruction. Learning-based deconvolution methods have also been explored, but they typically rely on paired training data, learned priors, or strong assumptions about the degradation model, which limits their applicability in single-image training-free scenarios\cite{weigert2018content,wang_deep_2019,qiao_neural_2025}.
	
	\subsection{Single-image self-supervised and zero-shot methods}
	
	Recent self-supervised and zero-shot denoising methods reduce the dependence on clean ground truth. Noise2Void\cite{krull2019noise2void} and Noise2Self\cite{batson2019noise2self} use blind-spot or masking strategies to learn denoising from noisy observations alone, but their effectiveness depends on assumptions about noise independence and spatial redundancy. In fluorescence microscopy, Noise2Fast\cite{lequyer2022fast}, SRDTrans\cite{li_spatial_2023}, ZS-DeconvNet\cite{qiao2024zero}, SN2N\cite{qu2024self}, and FAST\cite{wang_real-time_2025} further demonstrate the potential of label-free denoising. Among them, ZS-DeconvNet\cite{qiao2024zero} incorporates the physical imaging model and can be applied to deconvolution restoration.
	
	Despite these advances, many existing self-supervised methods still depend on masking schemes, data redundancy, repeated observations. Robust single-image restoration that simultaneously suppresses noise and performs deconvolution remains challenging.
	
	\subsection{Deep Image Prior}
	
	DIP provides a zero-shot restoration paradigm by optimizing a randomly initialized convolutional network for each degraded image individually\cite{ulyanov2018deep}. Since no external training data are required, DIP is attractive for fluorescence microscopy restoration where clean references are difficult to obtain. However, standard DIP may gradually overfit noise and artifacts during optimization, especially under low-SNR conditions. To mitigate this problem, previous studies have introduced strategies such as total variation regularization\cite{liu2019image}, Bayesian inference\cite{cheng2019bayesian}, RED regularization\cite{mataev2019deepred}, and early stopping\cite{li2021self},\cite{wang2021early}. More recent variants, including aSeqDIP\cite{alkhouri2024image} and Self-reinforcement DIP\cite{shu_sdip_2025}, improve DIP from the perspectives of sequential autoencoding regularization and input-adaptive optimization.
	
	Another important issue is that the effectiveness of the implicit prior of DIP may depend on the specific inverse problem. Although the convolutional network prior is effective in favoring coherent image structures in many restoration tasks, it may not always provide a sufficiently appropriate reconstruction bias for fluorescence microscopy deconvolution. In this task, the recovery of high-frequency details is highly ill-posed, and direct DIP-based deconvolution may produce unstable pseudo-structures or artifact-like details when the implicit prior alone is insufficient to constrain the solution space. Although DIP-related methods have been explored in optical microscopy, such as DRE-SIM for resolution extension of reconstructed SIM images\cite{he_surpassing_2023}, DIP-based zero-shot denoising and deconvolution for wide-field fluorescence microscopy images have not been systematically investigated. In this work, we address these issues with a stage-wise DIP-based restoration framework that first suppresses noise using aSeqDIP and then performs RL-guided DIP deconvolution. In the deconvolution stage, the RL result is used as a soft physical regularization prior to stabilize high-frequency recovery and mitigate artifact-prone solutions.

	\section{THE PROPOSED METHOD}
	\subsection{Problem Formulation and Overview}
	
	\begin{figure}[!ht]
		\centering
		\includegraphics[width=0.8\columnwidth]{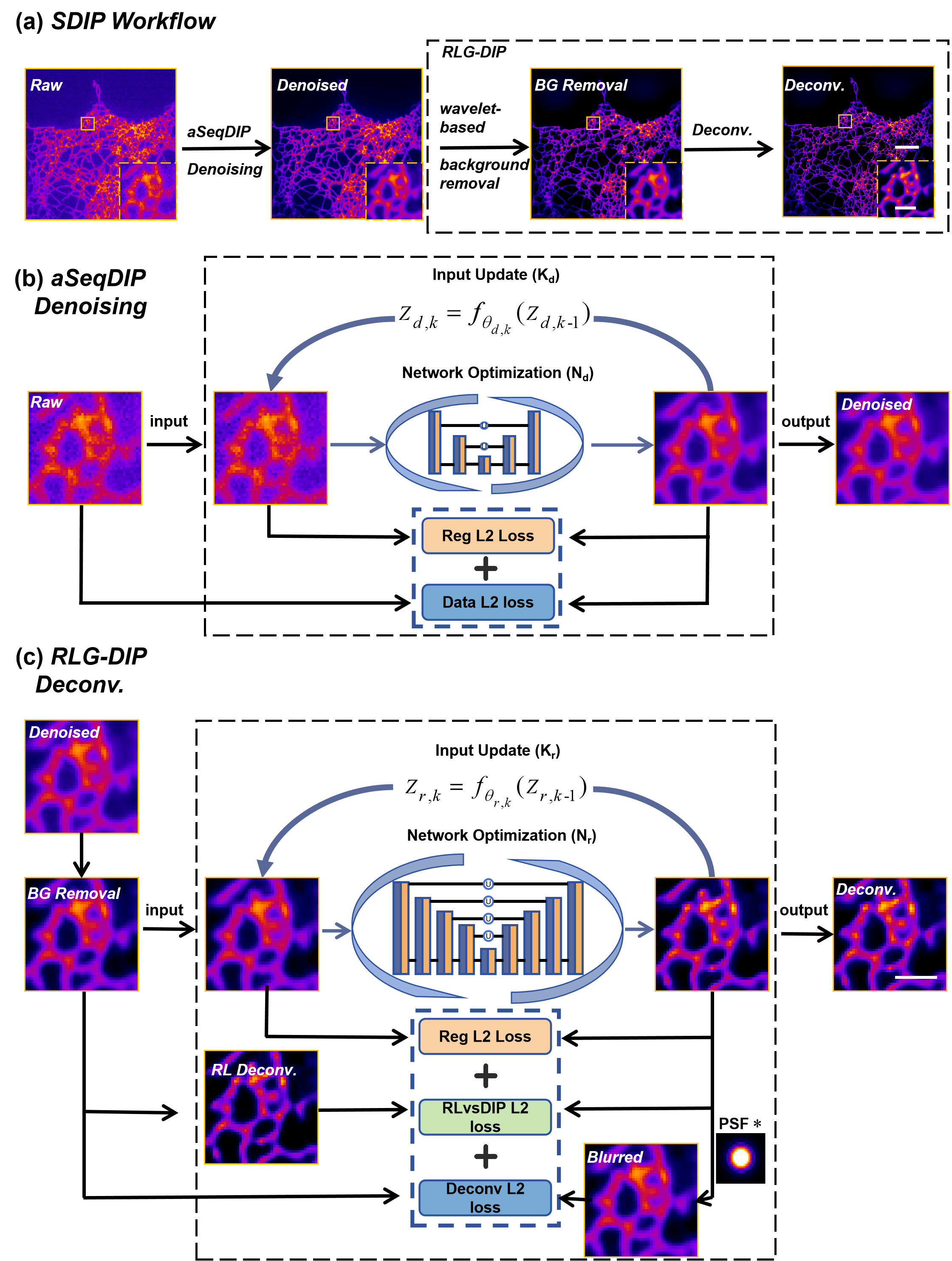}
		\caption{The proposed SDIP zero-shot framework for fluorescence microscopy image processing. (a) Stage-wise denoising and deconvolution pipeline for fluorescence microscopy images based on DIP. The raw noisy image is denoised by aSeqDIP, then fed into the RLG‑DIP module, where wavelet decomposition is first applied to remove background, followed by deconvolution for restoration. (b) aSeqDIP denoising module. It employs sequential autoencoding DIP optimization and recursively updates input and parameters to suppress noise and enhance structures. (c) RLG‑DIP deconvolution module. It takes the RL deconvolution result as a fixed prior, and constructs a joint loss under the constraint of PSF to improve detail restoration and resolution.Scale bar, 5 $\mu$m (a), 1 $\mu$m (magnified regions of a, b, c).}
		\label{fig1}
	\end{figure}

	Fluorescence microscopy images are often degraded by a combination of photon-limited noise, out-of-focus background, and diffraction-induced blur \cite{bertero_introduction_2021,zhao2022sparse}. 
	Given a single degraded observation \(y\), our goal is to recover a restored image \(\hat{x}\) with reduced noise, suppressed background interference, and enhanced structural resolution. The degradation process is modeled as
	\begin{equation}
		y = h * x + b + n,
		\label{eq:degradation_model}
	\end{equation}
	where \(x\) is the ground truth (GT), \(h\) denotes the PSF, \(*\) represents convolution, \(b\) is the background component, and \(n\) denotes measurement noise.
	
	Direct deconvolution of \(y\) can amplify noise and background, producing artifacts and unreliable structural details~\cite{bertero_introduction_2021}. Therefore, we adopt a stage-wise restoration strategy, as shown in Fig.~\ref{fig1}, in which denoising is performed before deconvolution:
	\begin{equation}
		y \rightarrow \hat{x}_d \rightarrow \hat{x},
		\label{eq:overall_pipeline}
	\end{equation}
	where \(\hat{x}_d\) is the denoised image and \(\hat{x}\) is the final restored result.
	
	Specifically, the input image \(y\) is first processed by an aSeqDIP-based denoising module to obtain \(\hat{x}_d\). The denoised image is then passed to the RLG-DIP deconvolution stage. Within this stage, a wavelet-based background correction is applied to remove background components before deconvolution. The corrected image is subsequently used as the input for RLG-DIP. The overall workflow is expressed as
	\begin{align}
		\hat{x}_d &= \mathcal{D}_{\mathrm{aSeqDIP}}(y), 
		\label{eq:aseqdip_denoising} \\
		\hat{x} &= \mathcal{R}_{\mathrm{RLG\mbox{-}DIP}}(\hat{x}_d, h),
		\label{eq:rlg_dip_deconvolution}
	\end{align}
	where \(\mathcal{D}_{\mathrm{aSeqDIP}}(\cdot)\) and \(\mathcal{R}_{\mathrm{RLG\mbox{-}DIP}}(\cdot)\) denote the aSeqDIP denoising process and the RLG-DIP deconvolution process, respectively. By performing deconvolution after denoising and background removal, the proposed stage-wise design reduces noise amplification while preserving fluorescence structures.

	\subsection{Deep Image Prior Preliminaries}
	DIP exploits the implicit regularization of an untrained convolutional neural network for image restoration. Instead of learning from an external training dataset, DIP optimizes the network parameters for each degraded image individually. Given a fixed random input \(z\), the restored image is parameterized as
	\begin{equation}
		x_{\theta} = f_{\theta}(z),
		\label{eq:dip_parameterization}
	\end{equation}
	where \(f_{\theta}(\cdot)\) denotes a convolutional neural network with parameters \(\theta\). For a generic restoration problem with observation \(y\), DIP estimates the restored image by solving
	\begin{equation}
		\theta^{*} = \arg\min_{\theta} \mathcal{L}\left(f_{\theta}(z), y\right),
		\label{eq:dip_general_loss}
	\end{equation}
	where \(\mathcal{L}(\cdot)\) is a task-dependent data fidelity term.
	
	For image denoising, the standard DIP objective is commonly written as
	\begin{equation}
		\theta^{*} = \arg\min_{\theta} 
		\left\| f_{\theta}(z) - y \right\|_2^2,
		\label{eq:dip_denoising_loss}
	\end{equation}
	and the restored image is obtained by
	\begin{equation}
		\hat{x} = f_{\theta^{*}}(z).
		\label{eq:dip_output}
	\end{equation}
	The effectiveness of DIP comes from the fact that convolutional networks tend to fit natural image structures earlier than random noise\cite{ulyanov2018deep}. However, when the optimization continues for too many iterations, the network may gradually overfit noise and artifacts in the observation\cite{wang2021early}. This limitation is particularly problematic for fluorescence microscopy images, where photon-limited noise and background can be further amplified in subsequent deconvolution.
	
	\subsection{aSeqDIP-Based Denoising}
	
	In the first stage, we employ aSeqDIP\cite{alkhouri2024image} to suppress noise in the degraded fluorescence image before deconvolution. Standard DIP optimizes an untrained network with a fixed random input, which may eventually overfit noise during prolonged optimization. In contrast, aSeqDIP introduces an input-adaptive sequential optimization scheme, where the network input is progressively updated by the network output, and an autoencoding regularization term is imposed to mitigate noise overfitting, as shown in Fig.\ref{fig1} (b).

	For the denoising task, we initialize the input as \( z_{d,0} = y \). Let \(f_{\theta_{d,k}}(\cdot)\) denote the denoising network at the \(k\)-th sequential step, where \(\theta_{d,k}\) represents its parameters, and the subscript \(d\) represents denoising. At each step, the network takes the previous input \(z_{d,k-1}\) and optimizes the following objective:
	\begin{equation}
		\begin{split}
			\theta_{d,k}^{*}
			=
			\arg\min_{\theta_{d,k}}
			&\left\| f_{\theta_{d,k}}(z_{d,k-1}) - y \right\|_2^2 \\
			&+
			\lambda_d^{\text{reg}} 
			\left\| f_{\theta_{d,k}}(z_{d,k-1}) - z_{d,k-1} \right\|_2^2,
		\end{split}
		\label{eq:aseqdip_objective}
	\end{equation}
	where the first term is the data fidelity loss (Data L2 loss) that penalizes deviation from the observed image \(y\), and the second term is the autoencoding regularization loss (Reg L2 loss) that enforces consistency between the network input and output. The regularization weight \(\lambda_d^{\text{reg}} \) controls the strength of the autoencoding constraint. 
	After optimizing the network parameters \(\theta_{d,k}\) for \(N_d\) steps, the network input is updated by :
	\begin{equation}
		z_{d,k}
		=
		f_{\theta_{d,k}}(z_{d,k-1}).
		\label{eq:aseqdip_update}
	\end{equation}
	The parameters \(\theta_{d,k}\) are initialized from the optimized parameters of the previous step, i.e., \(\theta_{d,k} \leftarrow \theta_{d,k-1}^{*}\) for \(k>1\), while the first step is initialized randomly as in standard DIP.
	
	After \(K_d\) sequential input updates, the final denoised image is obtained as
	\begin{equation}
		\hat{x}_d =z_{d,K_d}.
		\label{eq:aseqdip_final_output}
	\end{equation}
	
	This formulation differs from standard DIP in two aspects. First, the network input is no longer fixed but is sequentially updated using the output from the previous step. Second, the autoencoding term regularizes the network by encouraging the output to remain consistent with its input, which delays noise overfitting while preserving coherent fluorescence structures.

	\subsection{RLG-DIP Deconvolution}
	
	After denoising, the image \(\hat{x}_d\) is passed to the RLG-DIP deconvolution stage. To reduce the influence of background, a wavelet-based\cite{zhao2022sparse} correction is first applied:
	\begin{equation}
		b = W^{-1}\left(\mathcal{L}\left(W(\hat{x}_d)\right)\right),
		\label{eq:wavelet_background}
	\end{equation}
	where \(W(\cdot)\) and \(W^{-1}(\cdot)\) denote the wavelet transform and its inverse, respectively, and \(\mathcal{L}(\cdot)\) retains the low-frequency components. The corrected observation is obtained as
	\begin{equation}
		u_0 = \hat{x}_d - b.
		\label{eq:background_corrected_observation}
	\end{equation}

	The proposed RLG-DIP deconvolution follows the same optimization framework as aSeqDIP. Different from the denoising stage, however, the deconvolution objective incorporates the imaging forward model and an RL-guided prior to promote physically consistent deblurring.
	
	Let \(f_{\theta_{r,k}}(\cdot)\) denote the deconvolution network at the \(k\)-th sequential step, where \(\theta_{r,k}\) represents its parameters, and the subscript \(r\) represents deconvolution. 
	At each step, the network takes the previous input \(z_{k-1}\) and optimizes the following objective:
	\begin{equation}
		\begin{split}
			\theta_{r,k}^{*}
			=
			\arg\min_{\theta_{r,k}}
			&\left\| h * f_{\theta_{r,k}}(z_{r,k-1}) - u_0 \right\|_2^2 \\
			&+
			\lambda_r^{\text{RL}}
			\left\| f_{\theta_{r,k}}(z_{r,k-1}) - x_{\mathrm{RL}} \right\|_2^2\\
            &+ \lambda_{r}^{\text{reg}} \left\| f_{\theta_{r,k}}(z_{r,k-1}) - z_{r,k-1} \right\|_2^2 ,
		\end{split}
		\label{eq:rlgdip_objective}
	\end{equation}

	Here, the first term is the deconvolution data consistency loss (Deconv L2 loss), which ensures that the deconvolved estimate, after being blurred by the same PSF, remains consistent with the background-corrected observation \(u_0\). The second term is the RL-guided regularization loss (RLvsDIP L2 loss), which encourages the network output to align with the RL deconvolution prior \(x_{\mathrm{RL}}\).The third term is the autoencoding regularization loss (Reg L2 loss) that enforces consistency between the network input and output. \(\lambda_{r}^{\text{RL}}\) and \(\lambda_{r}^{\text{reg}}\)  are the corresponding regularization weights.
    
	\(x_{\mathrm{RL}}\) is calculated from the RL update\cite{richardson1972bayesian,lucy1974iterative},
	\begin{equation}
		x_{\mathrm{RL}}^{(j+1)}
		=
		x_{\mathrm{RL}}^{(j)}
		\cdot
		\left[
		h^{\top} *
		\frac{u_0}{h * x_{\mathrm{RL}}^{(j)} + \epsilon}
		\right],
		\label{eq:rl_update}
	\end{equation}
	where \(h^{\top}\) is the flipped PSF and \(\epsilon\) is a small constant used for numerical stability. Here, \(x_{\mathrm{RL}} = x_{\mathrm{RL}}^{(N_{\mathrm{RL}})}\) denotes the RL deconvolution result after \(N_{\mathrm{RL}}\) iterations.
	
	After optimizing \(\theta_{r,k}\) for \(N_r\) steps, the sequential input is updated by
	\begin{equation}
		z_{r,k}
		=
		f_{\theta_{r,k}}(z_{r,k-1}).
		\label{eq:rlgdip_input_update}
	\end{equation}
	After \(K_r\) sequential steps, the final restored image is obtained as
	\begin{equation}
		\hat{x}_r = z_{r,K_r}.
		\label{eq:rlgdip_final_output}
	\end{equation}
	
	By integrating aSeqDIP-based denoising, wavelet-based background suppression, and RLG-DIP deconvolution into a unified stage-wise framework, the zero-shot method achieves a progressive optimization process from noise suppression to blur compensation, thereby effectively improving the quality of fluorescence microscopy images.
	
	\subsection{Implementation Details}
	The proposed method operates in a \emph{zero-shot} manner by directly optimizing each target image without requiring additional training data or pre-trained models. In the denoising stage, a two-level encoder-decoder aSeqDIP network was adopted, whereas in the deconvolution stage, a four-level encoder-decoder DIP network was employed. For optimization, the Adam optimizer was used with a learning rate of $1\times10^{-4}$. The sequential updating process of both denoising and deconvolution was performed for \(N_d=1000,N_r=2000\) epochs, with \(K_d=K_r=10\) network parameter updates carried out within each epoch.  All regularization weights \(\lambda\) (i.e., \(\lambda_{d,k}^{\mathrm{reg}}\), \(\lambda_{r,k}^{\mathrm{reg}}\), and \(\lambda_{r,k}^{\mathrm{RL}}\)) were set to 1. In RLG‑DIP, the background removal adopts a wavelet‑based estimation using a 6th‑order Daubechies wavelet as the basis, with a decomposition level of 7, where only the lowest‑frequency approximation coefficients at the deepest level  are retained while all high‑frequency detail coefficients (levels 1–7, including horizontal, vertical, and diagonal components) are set to zero; this estimation is iterated 3 times to ensure convergence, and it takes \(N_{RL}=10\) iters in Eq.\ref{eq:rl_update} to calculate the RL guidance. More details are given in code (\codeURL).

	\section{EXPERIMENTAL RESULTS}
	
	\subsection{Experimental Setting}
	
	The experiments in this study were conducted using the \emph{BioSR}\cite{qiao2021evaluation} dataset. This dataset provides raw fluorescence image sequences acquired during structured illumination microscopy. For each sample, raw images are recorded under three illumination directions $(0^\circ, 60^\circ, 120^\circ)$, and under each direction, three phase-shifted images $(0^\circ, 120^\circ, 240^\circ)$ are captured. The 9 raw images can be averaged to generate a single wide-field image, which was used as the input for the subsequent denoising and deconvolution modules.
	
	In addition, the dataset provides images at different noise levels for each sample. In this study, images from \emph{level 01} (the lowest SNR) to \emph{level 05} were used as the input images for the denoising module. For denoising evaluation, the image with the highest SNR in the corresponding dataset was used as the reference image to calculate the PSNR and SSIM of the denoising results. For deconvolution evaluation, the corresponding SIM reconstructed image was used as the reference image to evaluate the final restoration performance. Since the original SIM reconstructed images have a size of $1004\times1004$ pixels, whereas the raw wide-field images are $502\times502$ pixels, the SIM reconstructed images were downsampled to $502\times502$ pixels using $2\times2$ pixel binning to ensure spatial consistency with the raw input images.

	The imaging parameters were kept consistent across all data types, where the numerical aperture of the detection objective was 1.3, the excitation wavelength was 488 nm, and the pixel size was approximately 62.6 nm. These parameters were further used as the conditions for PSF generation in the deconvolution stage.

	\subsection{Denoising Results}
	
	\begin{figure}[!ht]
		\centerline{\includegraphics[width=\columnwidth]{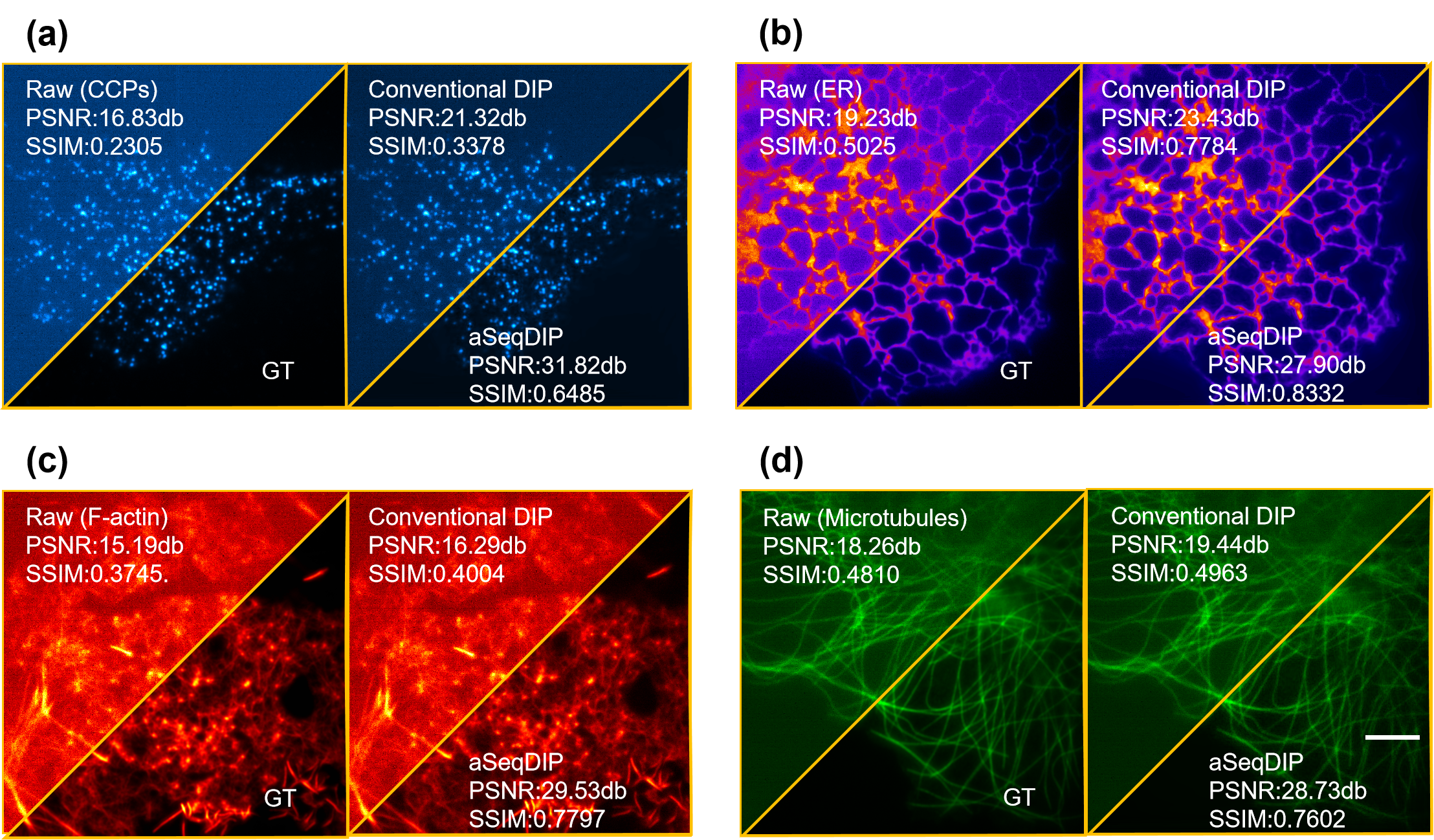}}
		\caption{Representative results of the aSeqDIP denoising module on samples with different cellular structures.(a–d) Denoising results on four typical subcellular structures (CCPs, ER, F-actin, Microtubules). For each sample, the upper-left region shows the input noisy image, and the lower-right region shows the corresponding clean ground-truth image. In the right panel, the upper-left is the denoising result of DIP, and the lower-right is the denoising result of aSeqDIP, with corresponding PSNR and SSIM values provided. Scale bar, 5 $\mu$m. }
		\label{fig2}
	\end{figure}
	
	\begin{table}[!htbp]
		\caption{Quantitative comparison results of the initial noisy image, DIP, and aSeqDIP for denoising under different noise levels. The corresponding PSNR and SSIM values are provided in the table.}
		\label{tab:denoise_psnr_ssim}
		\centering
		\renewcommand{\arraystretch}{1.15}
		\setlength{\tabcolsep}{8pt}
		\begin{tabular}{lccc}
			\toprule
			Methods& Initial & DIP & aSeqDIP \\
			\cmidrule(lr){2-4}
			& \multicolumn{3}{c}{PSNR(dB)/SSIM} \\
			\midrule
			level 01 & 17.60/0.3580 & 18.86/0.3979 & 26.48/0.6415 \\
			level 02 & 23.49/0.5562 & 25.08/0.5982 & 32.25/0.8114 \\
			level 03 & 28.99/0.7250 & 29.79/0.7463 & 35.37/0.9274 \\
			level 04 & 32.62/0.8366 & 32.93/0.8454 & 36.58/0.9512 \\
			level 05 & 35.30/0.9005 & 35.35/0.9008 & 37.75/0.9615 \\
			\bottomrule
		\end{tabular}
	\end{table}
	
	To evaluate the denoising performance of the  aSeqDIP relative to conventional DIP, Fig.\ref{fig2} presents visual comparisons on fluorescence microscopy samples with different cellular structures. Compared with the noisy inputs, both aSeqDIP and conventional DIP reduce the noise to some extent. However, aSeqDIP consistently yields more effective improvements, producing cleaner backgrounds, clearer structural boundaries, and better overall structural preservation, whereas conventional DIP provides only limited enhancement. This trend is observed across all four representative structural types shown in Fig.\ref{fig2}. One possible reason is that conventional DIP tends to gradually fit image noise during optimization, which may limit its final denoising performance. Overall, the visual comparisons indicate that aSeqDIP provides more effective and stable denoising performance.

	The quantitative results in Table~\ref{tab:denoise_psnr_ssim} further support the above visual observations. As the noise level gradually decreases from \emph{level 01} to \emph{level 05}, the PSNR and SSIM values of all three methods increase accordingly. More importantly, aSeqDIP consistently achieves the best PSNR and SSIM under all noise conditions, demonstrating its robustness across different SNR levels. For example, under \emph{level 01}, aSeqDIP achieves (26.48 dB / 0.6415), significantly outperforming the initial noisy image (17.60 dB / 0.3580) and conventional DIP (18.86 dB / 0.3979). Under \emph{level 05}, aSeqDIP still obtains the best result, reaching 37.75 dB / 0.9615, compared with (35.35 dB / 0.9008) for DIP and (35.30 dB / 0.9005) for the initial image. These quantitative results are consistent with the visual comparisons and suggest that the autoencoding strategy of aSeqDIP helps mitigate noise fitting during optimization while better preserving structural information.

	\subsection{RLG-DIP for deconvolution}
	\begin{figure}[!ht]
		\centerline{\includegraphics[width=\columnwidth]{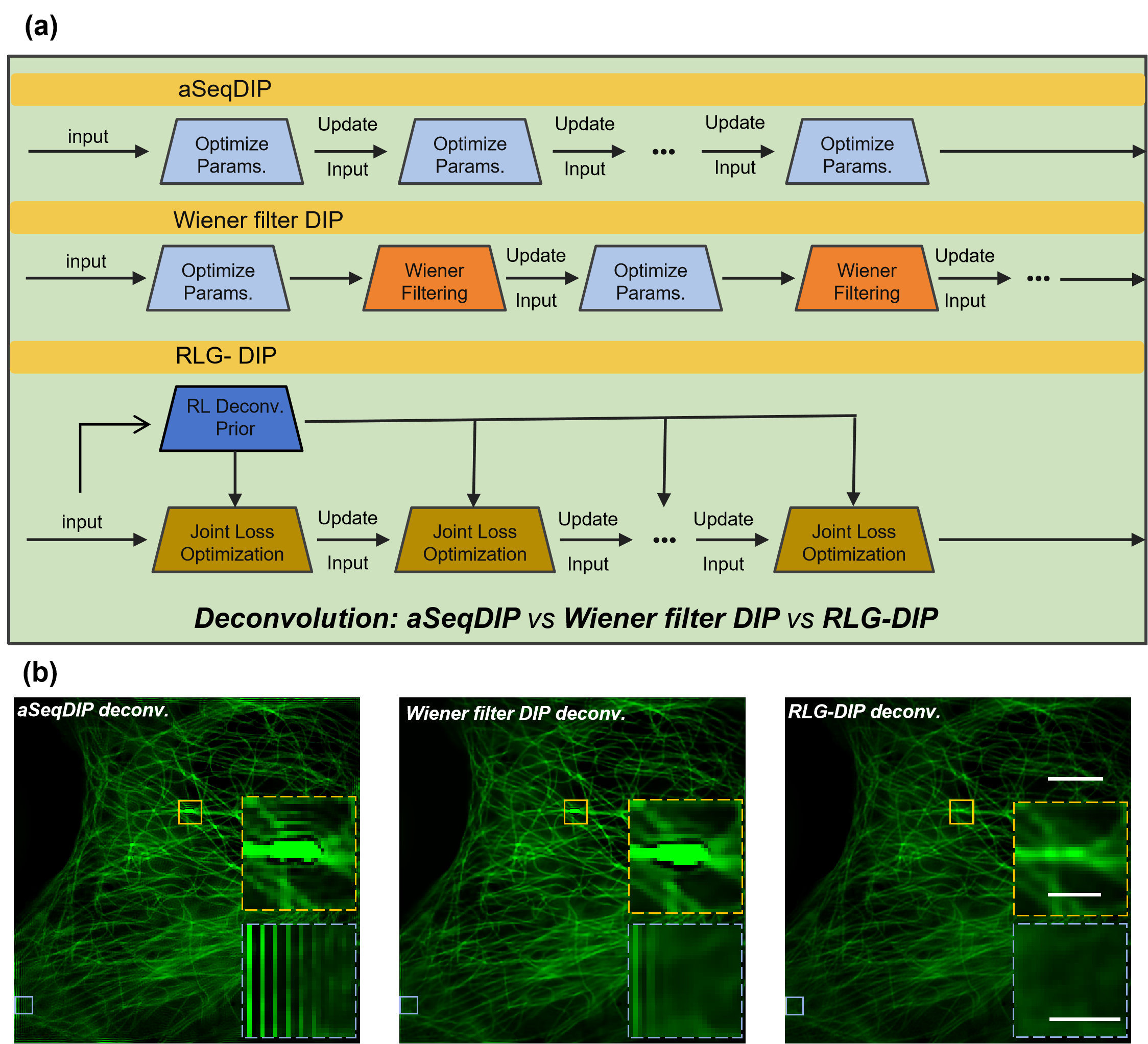}}
		\caption{Comparison of workflows and performance of different deconvolution DIP methods. 
			(a) Schematic of three deconvolution DIP workflows. 
			(b) Deconvolution results on microtubule structures. Magnified regions show the stripe artifact suppression effect of each method, where RLG-DIP effectively eliminates artifacts and recovers clear fibrous structures. 
			Scale bars: 5 $\mu$m (full view in (b)), 1 $\mu$m (yellow magnified boxes of b, blue magnified boxes of b).}
		\label{fig3}
	\end{figure}
	
	Directly applying aSeqDIP to deconvolution leads to noticeable artifacts and unstable pseudo-details, as shown in Fig. \ref{fig3}(b). Although aSeqDIP is effective at suppressing random noise and stabilizing coarse structures, it does not provide sufficiently reliable guidance for recovering high-frequency details during deconvolution. A possible reason is that deconvolution is a highly ill-posed inverse problem \cite{bertero_introduction_2021}. In general, obtaining a physically meaningful solution requires appropriate prior constraints, such as total variation, Hessian-based regularization \cite{huang_fast_2018}, or sparsity priors \cite{zhao2022sparse}. From this perspective, the regularization term restricts the solution space and determines the reconstruction bias. While the implicit prior of DIP is beneficial for denoising, it may not be sufficiently suitable for deconvolution, where high-frequency recovery is much more sensitive to prior mismatch. As a result, the optimization may be driven toward incorrect high-frequency patterns, leading to artifact amplification and unstable pseudo-details.
	
	Based on this intuition, we test two strategies to alleviate this problem. The first is Wiener-DIP, in which Wiener filtering is inserted into the sequential update process to suppress unstable high-frequency components before the next-stage update. However, stripe-like artifacts still remain in the reconstructed results, suggesting that the problem cannot be resolved by signal-level correction alone, but rather stems from the inadequacy of the DIP prior for deconvolution. The second is the proposed RLG-DIP, which introduces the RL deconvolution result as an explicit guidance prior during optimization. Unlike Wiener filtering, RL deconvolution is derived from the microscopy imaging model and explicitly accounts for both the PSF and Poisson noise statistics, thereby providing a more physically consistent reconstruction. By incorporating RL guidance into DIP optimization, RLG-DIP combines the implicit regularization of DIP with the physical modeling of RL, thereby suppressing artifacts and stabilizing fine-detail reconstruction, as shown in Fig.\ref{fig3}. This design choice is further validated by the experimental comparisons in later sections and Fig. \ref{fig4}.

	\subsection{Deconvolution Results}
	
	\begin{figure}[!ht]
		\centerline{\includegraphics[width=\columnwidth]{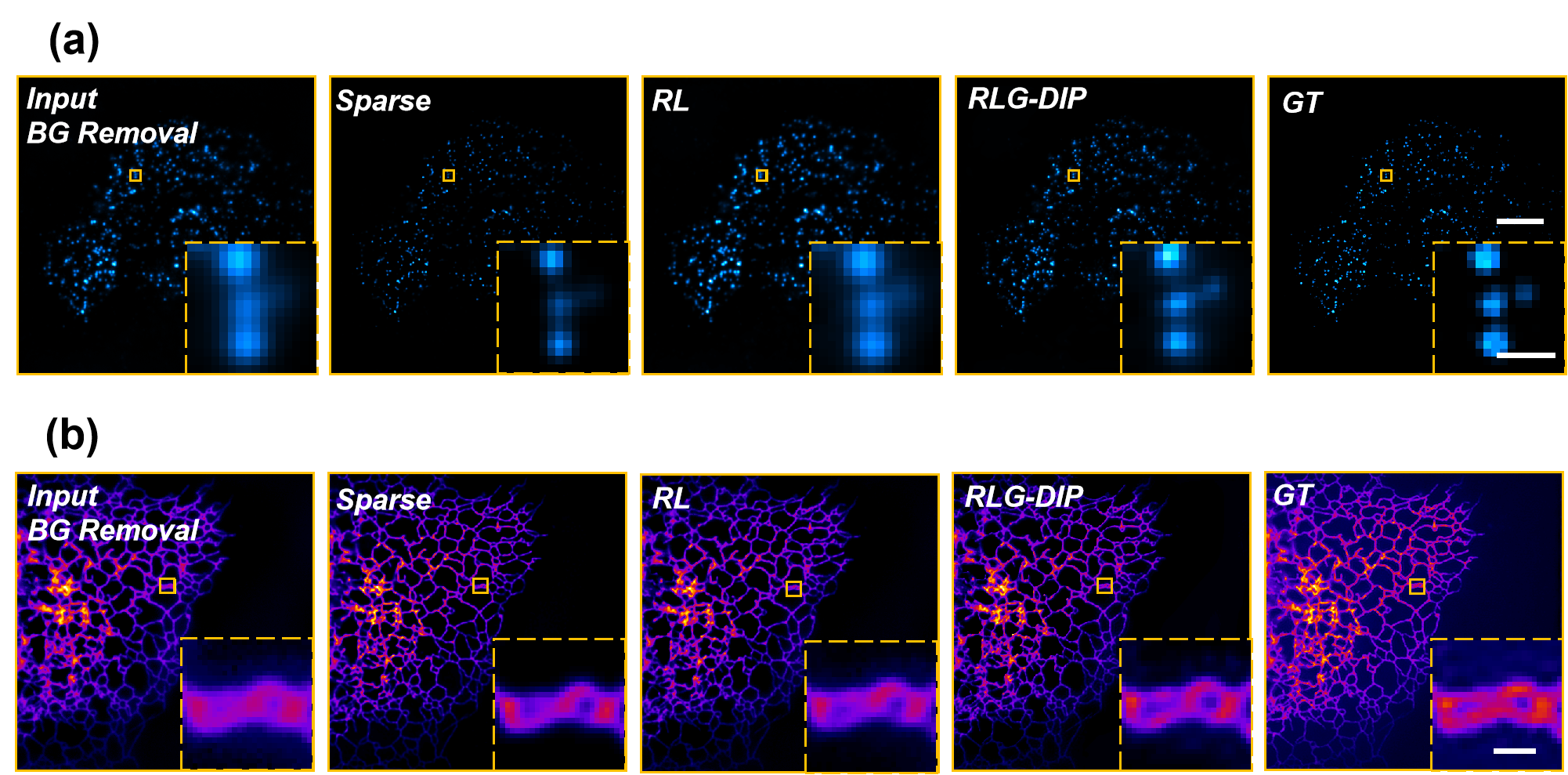}}
		\caption{Comparison of results by different deconvolution methods
			on representative cellular structure samples. The figure compares the restoration performance of the input image, the Sparse deconvolution method, the RL deconvolution method, the RL-guided DIP, and the ground-truth image on CCPs (a) and ER (b). A corresponding locally magnified region is provided in each panel to reveal the differences in detail restoration. Scale bars: 5 $\mu$m (full view), 0.5 $\mu$m (magnified regions of a, b).}
		\label{fig4}
	\end{figure}
	
	\begin{table}[!htbp]
		\caption{Quantitative comparison of Sparse, RL, and RLG-DIP for deconvolution on various cellular structures. The table lists corresponding PSNR and SSIM values.}
		\label{tab:deconv_psnr_ssim}
		\centering
		\renewcommand{\arraystretch}{1.15}
		\setlength{\tabcolsep}{8pt}
		\begin{tabular}{lccc}
			\toprule
			Methods& Sparse & RL & RLG-DIP\\
			\cmidrule(lr){2-4}
			& \multicolumn{3}{c}{PSNR(dB)/SSIM} \\
			\midrule
			CCPs & 33.94/0.9484 & 28.30/0.7925 & 31.74/0.8536 \\
			ER & 22.05/0.5658 & 27.47/0.6487 & 29.48/0.9059 \\
			F-actin & 22.92/0.4204 & 27.65/0.7489 & 29.97/0.8410 \\
			Microtubules & 22.99/0.4612 & 26.73/0.7886 & 29.35/0.8519 \\
			\bottomrule
		\end{tabular}
	\end{table}

	To evaluate the effectiveness of the proposed RLG-DIP deconvolution module, we compared it with conventional RL and Sparse deconvolution on representative cellular structures. Since this experiment focuses on the deconvolution module rather than the denoising stage, the raw wide-field images with the highest SNR in the corresponding BioSR dataset were used as the input images. This setting reduces the influence of severe noise and allows a more direct evaluation of deconvolution performance. For a fair comparison, all methods were applied to the same background-corrected observations, and the corresponding SIM reconstructed images were used as reference images for quantitative evaluation.
	
	As shown in the magnified regions of Fig.\ref{fig4}, the RL method recovers sharper high-frequency features than the input image, but it also introduces locally broadened intensity profiles around fine structures. The Sparse method improves foreground contrast and suppresses part of the background interference; however, some fine structures appear less continuous, and weak structural components are not fully preserved. In contrast, the proposed RLG-DIP module produces more visually stable restoration results, with clearer structural continuity, better preserved boundaries, and fewer visible deconvolution artifacts in the magnified regions. These observations suggest that incorporating RL-based physical guidance into DIP optimization helps stabilize high-frequency recovery while suppressing deconvolution artifacts.

	The quantitative results in Table~\ref{tab:deconv_psnr_ssim} further support the visual comparisons. RLG-DIP achieves the best PSNR and SSIM values on ER, F-actin, and microtubules. Specifically, on the ER sample, RLG-DIP obtains a PSNR/SSIM of 29.48 dB / 0.9059, outperforming Sparse deconvolution 22.05 dB / 0.5658 and RL deconvolution 27.47 dB / 0.6487. Similar improvements are observed on F-actin, where RLG-DIP reaches 29.97 dB / 0.8410, and on microtubules, where it achieves 29.35 dB / 0.8519. These results indicate that the RL-guided soft regularization can improve structural similarity and reconstruction accuracy for most evaluated structures.
	
	It should also be noted that RLG-DIP does not achieve the best quantitative performance on the CCPs sample. For CCPs, Sparse deconvolution obtains the highest PSNR/SSIM of 33.94 dB / 0.9484, whereas RLG-DIP achieves 31.74 dB / 0.8536. This may be because CCPs exhibit relatively sparse and compact structures, for which sparsity-based priors are particularly effective. Nevertheless, RLG-DIP outperforms both RL and Sparse deconvolution on most other structural types and shows more consistent visual restoration performance. Overall, these results demonstrate that the proposed RLG-DIP module provides an effective deconvolution strategy for fluorescence microscopy images by balancing RL-guided physical regularization and the implicit prior of DIP.

	\subsection{SDIP Restoration Results}
	
	\begin{figure}[!ht]
		\centerline{\includegraphics[width=\columnwidth]{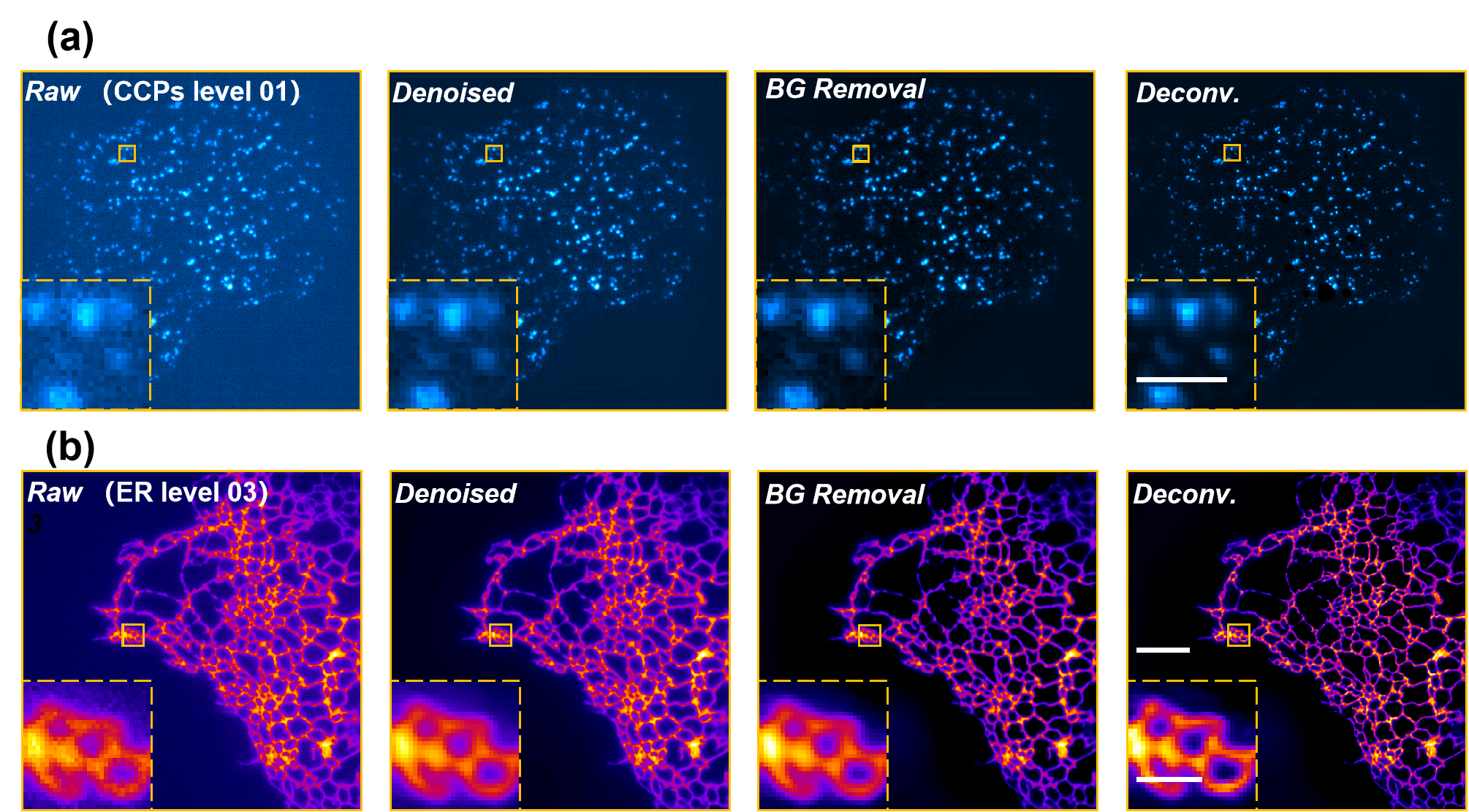}}
		\caption{Stage-wise enhancement results of the proposed full-flow SDIP method on samples with CCPs (a) and ER (b). The figure presents the visualization results including the noisy image, denoised image, background-removed image, and final deconvolved image, with corresponding locally magnified regions provided. scale bars: 5 $\mu$m (full view), 1 $\mu$m (magnified regions of a, b).}
		
		\label{fig5}
	\end{figure}
    
	\begin{table}[!htbp]
		\caption{Quantitative comparison of RL deconvolution, Sparse deconvolution, and SDIP on various cellular structures. The table lists corresponding PSNR and SSIM values averaged over five noise levels (\emph{level 01}  to \emph{level 05} ) for each cellular structure.}
		\label{tab:SDIP_psnr_ssim}
		\centering
		\renewcommand{\arraystretch}{1.15}
		\setlength{\tabcolsep}{8pt}
		\begin{tabular}{lccc}
			\toprule
			Methods& RL & Sparse & SDIP\\
			\cmidrule(lr){2-4}
			& \multicolumn{3}{c}{PSNR(dB)/SSIM} \\
			\midrule
			CCPs & 24.18/0.3763 & 33.40/0.7926 & 29.98/0.6859 \\
			ER & 20.62/0.6485 & 22.26/0.4272 & 24.59/0.7123 \\
			F-actin & 21.15/0.6738 & 23.53/0.4847 & 28.73/0.7702 \\
			Microtubules & 19.91/0.6120 & 23.74/0.5427 & 28.23/0.7901 \\
			\bottomrule
		\end{tabular}
	\end{table}

	To evaluate the overall effectiveness of the proposed SDIP framework, we applied the complete restoration pipeline to fluorescence microscopy images with different cellular structures. Different from the module-level evaluation in the previous subsection, this experiment assesses the final restoration performance of the complete stage-wise pipeline.
	
	As shown in Fig.\ref{fig5}, SDIP progressively improves the image quality through different restoration stages. The noisy input images contain obvious random noise, background interference, and blurred fine structures. After the denoising stage, random noise is effectively suppressed while the main cellular structures are largely preserved. The subsequent background correction step further reduces low-frequency background interference, which benefit further deconvolution. Finally, the RLG-DIP deconvolution stage enhances fine structural details and improves the separation of adjacent structures. In the magnified regions, SDIP produces clearer boundaries, better structural continuity, and more distinguishable fine details for both compact spot-like structures and continuous filamentous or reticular structures.
	
	The quantitative results in Table~\ref{tab:SDIP_psnr_ssim} further demonstrate the advantage of the full SDIP framework.These values are averaged over five noise levels (\emph{level 01} to \emph{level 05}) for each cellular structure from the BioSR dataset. Compared with RL and Sparse deconvolution, SDIP achieves the best PSNR and SSIM values on ER, F-actin, and microtubules. Specifically, SDIP obtains 24.59 dB / 0.7123 on ER, 28.73 dB / 0.7702 on F-actin, and 28.23 dB / 0.7901 on microtubules, outperforming both comparison methods on these structures. These results suggest that the complete SDIP pipeline can improve reconstruction accuracy and structural similarity for most evaluated cellular structures.
	
	For CCPs, Sparse deconvolution achieves the highest PSNR and SSIM values, with 33.40 dB / 0.7926, whereas SDIP obtains 29.98 dB / 0.6859. This result is consistent with the deconvolution module evaluation and suggests that sparsity-based priors may be particularly effective for compact and sparsely distributed structures. Nevertheless, SDIP shows more favorable performance on the other three structural types, indicating better overall adaptability across different fluorescence microscopy structures.

    \begin{table}[!htbp]
		\caption{Quantitative resolution comparison (mean resolution in nm) of different methods on the BioSR dataset.}
		\label{tab:resolution_compare}
		\centering
		\renewcommand{\arraystretch}{1.15}
		\setlength{\tabcolsep}{8pt} 
		\begin{tabular}{lccccc}
			\toprule
			Dataset & Raw & RL & Sparse & SDIP &  GT \\
			\cmidrule(lr){2-6}
			& \multicolumn{5}{c}{Resolution (nm)} \\
			\midrule
			CCPs       & 257.51& 204.48 & 145.45 & 147.27 & 137.16 \\
			ER         & 310.55 & 234.46 & 154.43 & 155.44 & 164.92 \\
			F-actin    & 302.06 & 255.99& 172.47 & 161.57 & 135.25 \\
			Microtubules&278.11 & 252.52 & 158.93 & 158.06 & 133.59 \\
			\bottomrule
		\end{tabular}
	\end{table}
	
	In addition to PSNR and SSIM, we further estimated the effective resolution of different restoration results using decorrelation analysis~\cite{descloux2019parameter}. As shown in Table~\ref{tab:resolution_compare}, all restoration methods improve the estimated resolution compared with the raw wide-field images, while SDIP achieves competitive or superior resolution performance on most structures. For F-actin and microtubules, SDIP obtains the best estimated resolution among the compared methods, reaching 161.57 nm and 158.06 nm, respectively. For ER, SDIP achieves 155.44 nm, which is comparable to Sparse deconvolution 154.43 nm. For CCPs, Sparse deconvolution shows a slight advantage, while SDIP still provides a substantial improvement over the raw and RL-restored images. These results further support the ability of SDIP to enhance fine structural details. It should be noted that decorrelation-based resolution estimation may be affected by contrast enhancement and high-frequency artifacts; therefore, the resolution results are interpreted together with visual comparisons and PSNR/SSIM evaluations rather than as the sole criterion for restoration quality.
	
	We also conducted an additional comparison with ZS-DeconvNet\cite{qiao2024zero}, a recent zero-shot fluorescence microscopy restoration method, on two representative BioSR samples. Under the same experimental setting, ZS-DeconvNet required more than six times the runtime of SDIP, making a large-scale comparison computationally expensive. Therefore, we report this comparison as a supplementary representative study, with detailed visual and quantitative results (see Appendix). SDIP achieves better stability and robustness than ZS‑DeconvNet while obtaining favorable metrics. ZS‑DeconvNet produces artifacts during training, whereas SDIP consistently maintains clear structural boundaries and continuous morphology. Although this comparison is limited to representative cases, it further supports the importance of explicit physical guidance for stabilizing zero-shot deconvolution.

	Overall, the full SDIP framework achieves consistent restoration improvement in terms of visual quality, PSNR/SSIM, and estimated resolution on most evaluated structures. The results also show that the performance of different restoration priors can depend on the structural characteristics of the biological sample. While sparse priors may be advantageous for compact spot-like structures such as CCPs, the proposed SDIP framework provides more stable and general restoration performance for filamentous and reticular structures.

	\section{Conclusion}
	In this work, we proposed SDIP, a zero-shot stage-wise restoration framework for fluorescence microscopy image denoising and deconvolution. SDIP first employs an aSeqDIP-based denoising module to suppress noise while preserving fluorescence structures, and then performs wavelet-based background correction followed by RLG-DIP deconvolution. In RLG-DIP, the Richardson--Lucy deconvolution result is introduced as a soft physical regularization prior, guiding DIP optimization toward physically plausible high-frequency recovery while reducing unstable deconvolution artifacts.
	
	Experiments on the BioSR dataset demonstrate that SDIP improves visual quality and achieves favorable quantitative performance on most evaluated cellular structures. Compared with RL and sparse deconvolution methods, SDIP shows better structural continuity, fewer visible artifacts, and improved PSNR/SSIM on ER, F-actin, and microtubules, while sparse priors remain advantageous for compact structures such as CCPs. These results suggest that combining the implicit prior of DIP with model-driven physical guidance is an effective strategy for training-free fluorescence microscopy restoration and may provide useful insights for physically guided DIP methods in other inverse problems.

    \section*{Acknowledgment}
    This work was supported by the Research Fund of Wenzhou Institute, UCAS (Grant No. WIUCASQD2021037 , WIUCASQD2021009 , WIUCAS2024007) and Wenzhou Key Laboratory Project (Grant No. SWGXXWCX). The authors would like to thank Prof. Xin Zhou for insightful discussions and for providing computational resources.

    \section*{Code Availability}
A PyTorch implementation of SDIP is available on GitHub at \codeURL.

    \section*{Competing Interests}
    The authors declare no competing interests.
    
	\section*{Appendix}
		\begin{figure}[!ht]
		\centerline{\includegraphics[width=\columnwidth]{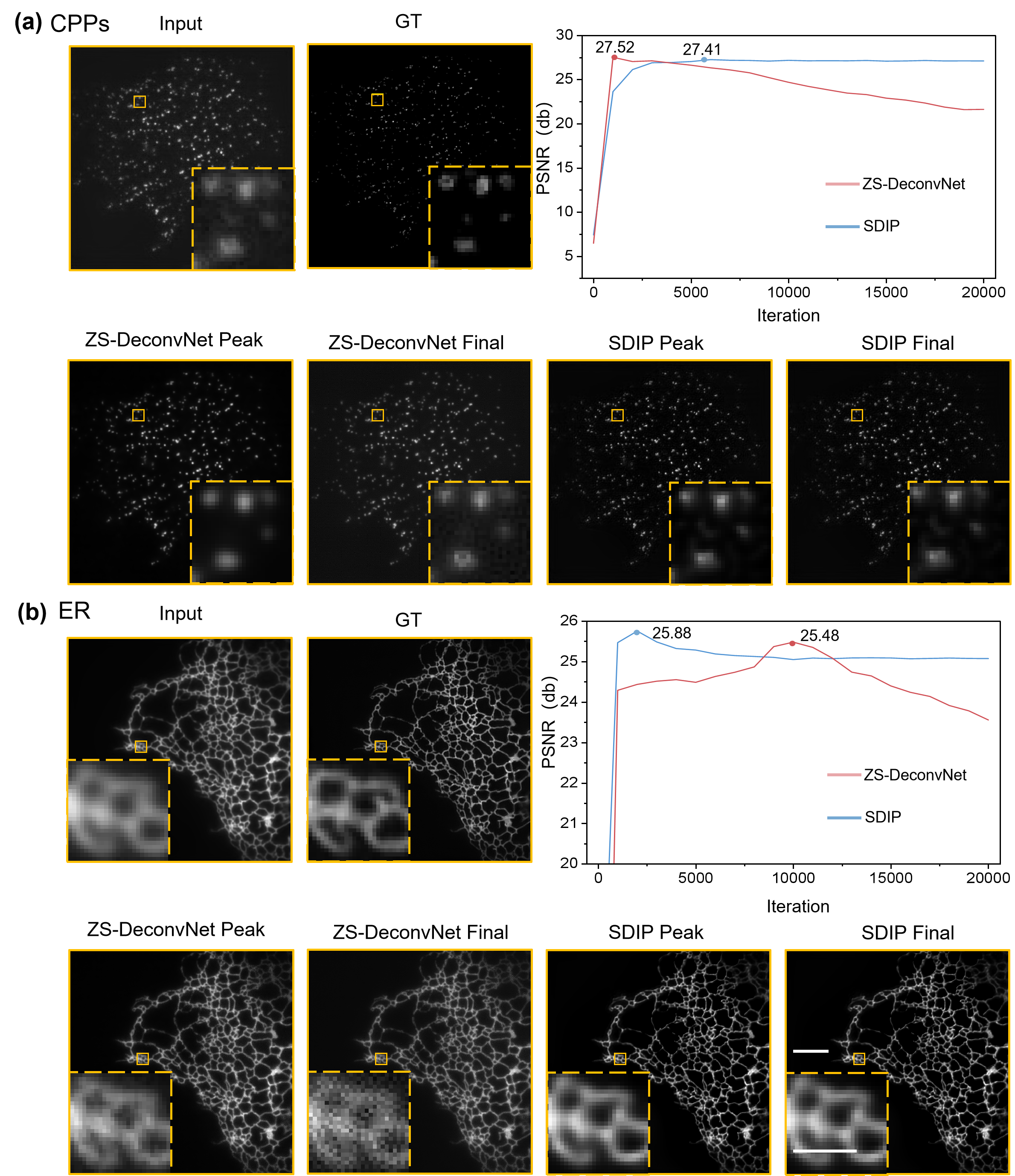}}
		\caption{Single‑image training curves and deconvolution results of SDIP versus ZS‑DeconvNet on the BioSR dataset.(a)CCPs structure, (b) ER structure. Each subpanel contains the input wide‑field image, the ground‑truth (SIM reconstruction) image, and the PSNR evolution curves of SDIP and ZS‑DeconvNet over iteration steps. The peak‑PSNR results of ZS‑DeconvNet and its final deconvolution output after 20,000 iterations are shown. The peak‑PSNR results of SDIP and its final output (2,000 input iterations, each with 10 network parameter updates, totaling 20,000 network iterations) are also shown. Scale bar: 5 $\mu$m (full view), 1 $\mu$m (magnified regions).}
		\label{fig6}
	\end{figure}
    In this appendix, we further compare the proposed zero-shot SDIP method with the recent state-of-the-art ZS‑DeconvNet \cite{qiao2024zero} on two representative BioSR samples (CCPs and ER). Under exactly the same experimental settings, we evaluate the reconstruction accuracy and optimization robustness of both methods using identical numbers of iterations and the same evaluation metrics. Fig.\ref{fig6} shows the single-image training curves and deconvolution results of SDIP and ZS‑DeconvNet on two typical cellular structures.
    
	As shown in Fig.\ref{fig6}(a) CCPs. In the training curves, the peak PSNR of SDIP (27.41 dB) is close to that of ZS‑DeconvNet (27.52dB). However, the SDIP curve remains stable after reaching its peak, whereas the ZS‑DeconvNet curve decreases noticeably over iterations, indicating that SDIP has a more stable optimization process. In the peak images, both methods recover most of the punctate structures. Nevertheless, in the final deconvolution outputs, ZS‑DeconvNet progressively fits noise, leading to a clear degradation in image quality, with visible background artifacts in the magnified regions. In contrast, SDIP preserves clear and coherent punctate structures in its final results.
    
    As shown in Fig.\ref{fig6}(b) ER. In the training curves, the peak PSNR of SDIP (25.88 dB) is higher than that of ZS‑DeconvNet (25.48 dB), and the PSNR of SDIP remains stable throughout the training process. In the peak images, SDIP achieves a cleaner background and sharper, more continuous structures compared to ZS‑DeconvNet. In the final deconvolution images, ZS‑DeconvNet produces abundant artifacts, while SDIP yields intact and clear tubular structures. The difference between the two methods is particularly evident in the magnified regions.
    
    The above comparisons demonstrate that SDIP not only achieves favorable PSNR metrics in zero‑shot deconvolution but also exhibits stronger robustness and reliability. In terms of visual quality, SDIP consistently maintains clear structural boundaries and continuous morphology.

	\bibliographystyle{IEEEtran}
	\bibliography{ref} 

\end{document}